\documentclass{article}
\usepackage{spconf,amsmath,graphicx}
\usepackage{bm}		
\usepackage{tikz}
\usepackage{pgfplots}
\usepackage{multirow}
\usepackage{rotating}
\usepackage{booktabs}
\usepackage{setspace}
\usepackage{makecell}
\usepackage{url}
\usepackage{hyperref}
\DeclareMathOperator*{\argmin}{arg\,min}

\def\taskLoss{\ensuremath{\mathcal{L}_\mathrm{task}}}
\def\rateLoss{\ensuremath{\mathcal{L}_\mathrm{rate}}}

\def\inputSignal{\ensuremath{\boldsymbol{x}}}
\def\inputSignalTransformed{\ensuremath{\boldsymbol{X}}}
\def\scalingMatrix{\ensuremath{\boldsymbol{S}}}
\def\quantizationStepSize{\ensuremath{\boldsymbol{\Delta}}}
\def\QP{\mathit{QP}}

\newcommand\tabrotate[1]{\begin{turn}{90}\rlap{#1}\end{turn}}	

\definecolor{crimson2143940}{RGB}{214,39,40}
\definecolor{darkorange25512714}{RGB}{255,127,14}
\definecolor{forestgreen4416044}{RGB}{44,160,44}
\definecolor{gainsboro217}{RGB}{217,217,217}
\definecolor{lightgray204}{RGB}{204,204,204}
\definecolor{steelblue31119180}{RGB}{31,119,180}
\newcommand{\showmark}[2]{
	\begin{tikzpicture}[baseline, baseline=-.5ex]	
		\node[#1, mark size=3, mark options={solid}]at (0mm,0){%
			\nullfont\pgfuseplotmark{#2}%
		};
	\end{tikzpicture}%
}

\newcommand\copyrighttext{%
	\footnotesize \textcopyright 2022 IEEE. Personal use of this material is permitted.
	Permission from IEEE must be obtained for all other uses, in any current or future 
	media, including reprinting/republishing this material for advertising or promotional 
	purposes, creating new collective works, for resale or redistribution to servers or 
	lists, or reuse of any copyrighted component of this work in other works. 
	DOI: \href{https://doi.org/10.1109/ICIP46576.2022.9897987}{ICIP46576.2022.9897987}}

\newcommand\copyrightnoticeOwn{%
	\begin{tikzpicture}[remember picture,overlay]
		\node[anchor=north,yshift=-10pt] at (current page.north) {\fbox{\parbox{\dimexpr\textwidth-\fboxsep-\fboxrule\relax}{\copyrighttext}}};
	\end{tikzpicture}%
	\vspace{-8mm}
}

\title{Learning Frequency-Specific Quantization Scaling in VVC for Standard-Compliant Task-driven Image Coding}
\name{Kristian Fischer, Fabian Brand, Christian Herglotz, and Andr\'e Kaup\thanks{The authors gratefully acknowledge that this work has been funded by the Deutsche Forschungsgemeinschaft (DFG, German Research Foundation) under project number 426084215.}}
\address{Multimedia Communications and Signal Processing\\
	Friedrich-Alexander-Universit\"at Erlangen-N\"urnberg (FAU)\\
	Cauerstr. 7, 91058 Erlangen, Germany\\
	\{Kristian.Fischer, Fabian.Brand, Christian.Herglotz, Andre.Kaup\}@fau.de}
\begin{document}
%
\maketitle
\copyrightnoticeOwn
\begin{abstract}
Today, visual data is often analyzed by a neural network without any human being involved, which demands for specialized codecs. For standard-compliant codec adaptations towards certain information sinks, HEVC or VVC provide the possibility of frequency-specific quantization with scaling lists.
This is a well-known method for the human visual system, where scaling lists are derived from psycho-visual models.
In this work, we employ scaling lists when performing VVC intra coding for neural networks as information sink. 
To this end, we propose a novel data-driven method to obtain optimal scaling lists for arbitrary neural networks.
Experiments with Mask R-CNN as information sink reveal that coding the Cityscapes dataset with the proposed scaling lists result in peak bitrate savings of 8.9\,\% over VVC with constant quantization.
By that, our approach also outperforms scaling lists optimized for the human visual system.
The generated scaling lists can be found under \url{https://github.com/FAU-LMS/VCM_scaling_lists}.	

\end{abstract}
\begin{keywords}
Video Coding for Machines, Scaling Lists, Adapted Quantization, VVC, Instance Segmentation
\end{keywords}
\vspace{-2mm}
\section{Introduction}
\vspace{-1mm}
Modern hybrid video codecs such as HEVC~\cite{sullivan2012_HEVC} and VVC~\cite{bross2021_VVC} allow for a frequency-specific quantization of the transform coefficients by providing scaling lists to the coder. In total, a scaling list in VVC consists of 28 scaling matrices for each prediction mode, blocksize, and color component~\cite{schwarz2021}. 
Scaling lists have mostly been employed to optimize the coding quality for the human visual system~(HVS). In 1999, Chang et~al.~\cite{chang1999} proposed JPEG quantization tables that are optimized for human perception modeled by a non-linear point transformation and a modulation transfer function. Later, the same model was adopted to derive the default quantization matrices for HEVC~\cite{sze2014_HEVC_Book, haque2011_HEVC_QM}. Prangnell et al.~\cite{prangnell2016} proposed quantization matrices for HEVC that are optimized for the human perception on high-resolution multimedia data. Those HVS-optimized matrices commonly exploit that the HVS is less sensitive to high spatial frequency in  visual content. Thus, high frequencies are quantized more coarsely to save bitrate while keeping a similar visual quality.

Due to the rapid progress in the field of neural networks solving tasks such as object detection, semantic segmentation, or tracking, the amount of coding scenarios where the visual data is directly analyzed by a neural network rather than being observed by a human is constantly increasing. 
This requires new, optimized coding schemes. As a result of this, MPEG introduced an ad-hoc group on \textit{video coding for machines~(VCM)} in 2019~\cite{zhang2019} that aims at standardizing efficient bitstreams for such machine-to-machine~(M2M) scenarios. Previous approaches to improve the coding efficiency of HEVC or VVC for VCM scenarios mainly consist of adding spatial saliency information to the encoding process~\cite{galteri2018, choi2018, fischer2021_ICASSP}. In addition, we proposed a feature-based rate distortion optimization for VVC in \cite{fischer2020_FRDO}. Optimizing scaling lists for algorithms as information sink has been proposed in~\cite{duan2012} and~\cite{chao2013} for JPEG. A broad analysis on the frequency sensitivity of image classification networks has been made in~\cite{tsuzuku2019} by adding perturbations based on Fourier basis functions.

Motivated by this development, this paper proposes a data-driven method to train optimal scaling matrices when coding images for an arbitrary neural network. Eventually, this scaling list is added to the VVC encoding process in order to reduce the amount of bits that is spent for frequency coefficients that are less important for the applied instance segmentation network Mask R-CNN~\cite{he2017}.

\begin{figure*}[!t]
	\centering
	\includegraphics[width=0.9\linewidth]{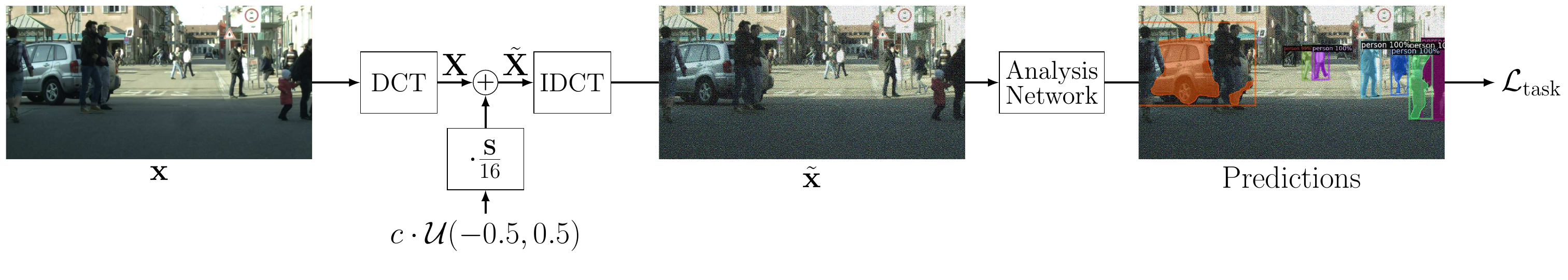}
	\vspace{-3mm}
	\caption{Proposed method to train the scaling matrix $\scalingMatrix$ for the DCT blocksize $B$. 
		$X$ and $\tilde{X}$ are of size $3 \times B^2 \times H/B \times W/B$. $\scalingMatrix$ is of size $B^2$ and added to each RGB color channel. 
		IDCT denotes the inverse DCT.
	}
	\vspace{-3mm}
	\label{fig:proposed framework}
\end{figure*}

\vspace{-1mm}
\section{Background}
\vspace{-1mm}
In hybrid video codecs, the error signal $\inputSignal$ of height $H$ and width $W$ between the prediction and the original signal is first transformed into $\inputSignalTransformed$ in order to condense the signal energy on few coefficients. To that end, a 2D frequency transform such as the discrete cosine transform~(DCT) is applied. Afterwards, $\inputSignalTransformed$ is uniformly quantized resulting in quantization indices $\hat{\inputSignalTransformed}$ that are eventually entropy coded and transmitted. At the decoder, the whole process is reverted and the transform coefficients $\hat{\inputSignalTransformed}$ are transformed back into the decoded signal $\hat{\inputSignal}$. The quantization step size $\Delta$ is derived from the user-defined quantization parameter~($\QP$) by
\begin{equation}
	\Delta = 2^{(\QP-4)/6}\cdot2^{\beta-8},
	\label{eq:qp to delta}
\end{equation}
with $\beta$ defining the bit depth of the signal $\inputSignal$ to ensure a similar quality for all bit depths~\cite{schwarz2021}.
Normally, each transform coefficient $\inputSignalTransformed_k$ at position $k$ is quantized with the same step size $\Delta$. However, HEVC and VVC also provide the possibility to define a scaling matrix $S$ that allows for a specific quantization of each frequency coefficient~\cite{schwarz2021, sze2014_HEVC_Book}. By that, the quantization of $\inputSignalTransformed_k$ changes to 
\begin{equation}
	\hat{\inputSignalTransformed}_k = \left\lfloor  \frac{\inputSignalTransformed_k}{\quantizationStepSize_k} + 0.5 \right\rfloor,
	\quad \mathrm{with} \quad
	\quantizationStepSize_k = \Delta \cdot \frac{\scalingMatrix_k}{16}.
	\label{eq:delta k}
\end{equation}
The coefficients in the scaling matrix are defined as positive integers, with $0\mkern-2mu<\mkern-2mu\scalingMatrix_k\mkern-2mu<\mkern-2mu16$ resulting in a smaller quantization interval and $\scalingMatrix_k\mkern-2mu>\mkern-2mu16$ resulting in a coarser quantization step size than the initial quantization step size $\Delta$ derived from $\QP$.


\section{Proposed Scaling List Generation}
\label{sec: Proposed Scaling List Generation}

\subsection{Scaling Matrix Generation}
To train the VCM-optimized scaling matrix $\scalingMatrix$, we build up the framework depicted in Fig.~\ref{fig:proposed framework}, which emulates the signal flow of the coding chain.
Contrary to the real hybrid coding chain, we applied three adaptations in order to make an end-to-end training possible. 
First, we utilized RGB images as input data $\inputSignal$ instead of error signals. 
Second, we focused on the DCT transformation, since it is the most important representative of the VVC transforms.
Third, quantization is not applicable to gradient-descent optimization due to its non-differentiable characteristic. Thus, we emulate the quantization by adding uniform noise similar to the field of learned end-to-end image compression~\cite{balle2017endtoend}. The strength of the noise is steered by the user-defined constant $c$. 
Hence, the distorted transform coefficients $\tilde{\inputSignalTransformed}$ are calculated by 
\begin{equation}
	\tilde{\inputSignalTransformed} = \inputSignalTransformed + \mathcal{U}(-0.5, 0.5) \cdot c \cdot \frac{\scalingMatrix}{16}.
\end{equation} 
With the trainable scaling matrix $\scalingMatrix$, the noise can separately be amplified for each frequency coefficient $\inputSignalTransformed_k$. Per design, we bound $\scalingMatrix$ between 16 and 128 by applying a Sigmoid non-linearity:
\begin{equation}
	\scalingMatrix = 16 + 112\cdot \mathrm{Sigmoid}(\scalingMatrix').
\end{equation}
Here, $\scalingMatrix'$ is the actual trainable parameter resulting in the bounded scaling matrix $\scalingMatrix$. The borders of 16 and 128 are inspired by the minimum and maximum values given for the default HEVC matrix~\cite{haque2011_HEVC_QM} and the work in~\cite{prangnell2016}. By that, we only allow the network to increase $\quantizationStepSize_k$. With the maximum $\scalingMatrix$ of 128, $\quantizationStepSize_k$ is limited to eight times the initial step size $\Delta$.

When training the scaling matrix, two contrary targets are pursued. The main goal of the proposed method is to increase $S$ in order to enlarge the quantization step size, and thus to ultimately reduce the bitrate during encoding in the inference case. To achieve this goal, we introduce a loss 
\begin{equation}
	\rateLoss(\scalingMatrix) = \frac{16}{\mathrm{mean}(\scalingMatrix)},
\end{equation}
approximating the likely rate savings later in inference and
forcing the network to increase the coefficients of $\scalingMatrix$. Due to the mean, the loss is independent of the DCT blocksize $B$. 

Naturally, only training on $\rateLoss$ would result in the maximum possible value of $\scalingMatrix$, neither considering the output quality resulting from $\scalingMatrix$ in inference nor weighting the different frequency coefficients. Therefore, we also consider the task loss $\taskLoss$ of the analysis network depending on the present noise in $\tilde{\inputSignal}$. With that, the network is pushed to increase the noise for the coefficients that are less harmful for the analysis network. The overall minimization problem is defined as
\begin{equation}
	\scalingMatrix = \argmin_{\scalingMatrix} \taskLoss(\tilde{\inputSignal}|\scalingMatrix) + \lambda \cdot \rateLoss(\scalingMatrix),
\end{equation}
where the parameter $\lambda$ weights the training towards one of the competing goals of a large scaling matrix, presumably resulting in a lower bitrate, or a high task performance.

\begin{figure}[!t]
	\centering
	\includegraphics[width=\linewidth]{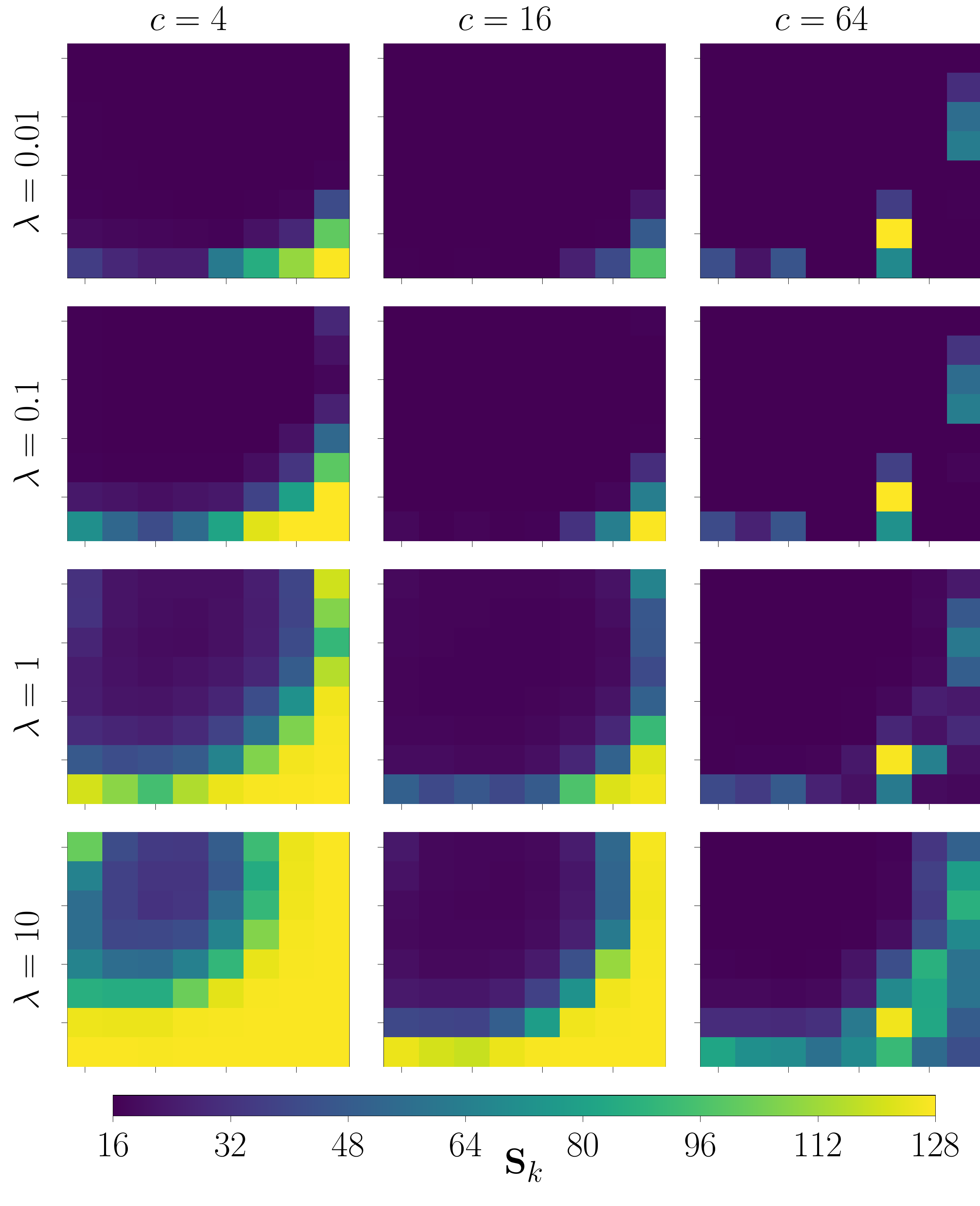}
	\vspace{-8mm}
	\caption{Resulting $8\times 8$ scaling matrices for different combinations of initial noise strength $c$ and loss weighting parameter $\lambda$. The direct component is located in the top left corner. A yellow value corresponds to a larger scaling matrix value.}
	\label{fig:resulting scaling lists}
	\vspace{-3mm}
\end{figure}

\subsection{Training Setup}
As analysis network in training, we employed the state-of-the-art instance segmentation network Mask R-CNN~\cite{he2017} with its proposed loss as $\taskLoss$. Its pre-trained weights were taken from the Detectron2 library~\cite{wu2019detectron2}.
We trained our framework on the 2965 RGB Cityscapes~\cite{cordts2016} training images, which were cropped to patches of $512\times 1024$ pixels. The scaling list was optimized by the Adam optimizer for 20 epochs. After 10 epochs, the initial learning rate of 0.01 was decreased to 0.001. The batch size was set to 16. With each training run, we trained one scaling matrix $\scalingMatrix$ for one DCT blocksize $B$, a given noise strength $c$, and one $\lambda$ value. The  scaling matrix was applied to each RGB color channel. Finally, the learned scaling matrix is rounded to the nearest integer.

\subsection{Resulting Scaling Matrices}

In Fig.~\ref{fig:resulting scaling lists}, the scaling matrices generated by our proposed framework are depicted for a DCT blocksize of $8\times 8$ and several combinations of $c$ and $\lambda$. The smaller $c$, and therewith the smaller the initial noise, the higher the network increases the values in the scaling matrix due to the lower impact on the task accuracy. For a high noise strength, only very few coefficients are increased, eventually resulting in only very few quantization step sizes $\quantizationStepSize_k$ to be enlarged, when applying the scaling matrix for coding. 

Fig.~\ref{fig:resulting scaling lists} also shows the influence of the loss weighting. For a low $\lambda$, the task loss has a high priority which results in the majority of scaling coefficients to be near the minimum value of 16 to avoid a noise amplification. With increasing $\lambda$, the rate loss $\rateLoss$ is more prioritized. Therefore, the training results in larger scaling coefficients, since it is more important to increase the scaling coefficients at the drawback of an increased noise leading to a higher task loss. To put it into a nutshell, the proposed method reveals that the evaluated Mask R-CNN model is less sensitive to deterioration in the high frequency components, which is in line with the findings in~\cite{tsuzuku2019}. 

\section{Performance Evaluation in VVC}
In this section, we demonstrate that the scaling lists derived from our measurement framework result in coding gains over conventional coding without frequency-adaptive quantization and over HVS-optimized scaling lists.
\subsection{Evaluation Setup}
For evaluation, we coded the 500 Cityscapes validation images with the standard-compliant VVC test model~(VTM)~\cite{chen2020vtm10} version 10.0 following the work in~\cite{fischer2020_ICIP} and the MPEG VCM CTCs~\cite{liu2020_VCM_CTC}. We selected $\QP$ values of 12, 17, 22, and 27 in order to obtain a high task accuracy close to the performance of uncompressed images, which would typically be required for practical applications. To measure the inference task performance, we utilized the weighted average precision~(wAP) metric as proposed in~\cite{fischer2020_ICIP}, which weights the state-of-the-art object detection metric, average precision, according to the class frequency to level class imbalances in the Cityscapes dataset. To quantify the coding gains of our proposed method, we calculate the Bj\o ntegaard delta rate~(BDR)~\cite{bjontegaard2001_new} with the wAP as quality metric. It measures the bitrate change compared to the anchor codec VTM-10.0 without scaling lists at the same accuracy of Mask R-CNN applied to the decoded images.

In order to obtain the scaling lists, we trained the scaling matrices with our proposed framework as described in Section~\ref{sec: Proposed Scaling List Generation} for squared blocks of size 2, 4, 8, 16, 32, and 64. 
The resulting scaling matrices measured in RGB colorspace were taken for the luma and the chroma channels respectively, and combined to one scaling list as described for VTM by JVET~\cite{toma2018_ScalingMatrices}. There, it is also described how to interpolate the scaling lists for rectangular blocks.

\begin{figure}[!t]
	\centering
	\includegraphics[trim=20 0 0 0mm, width=0.95\linewidth, clip=true]{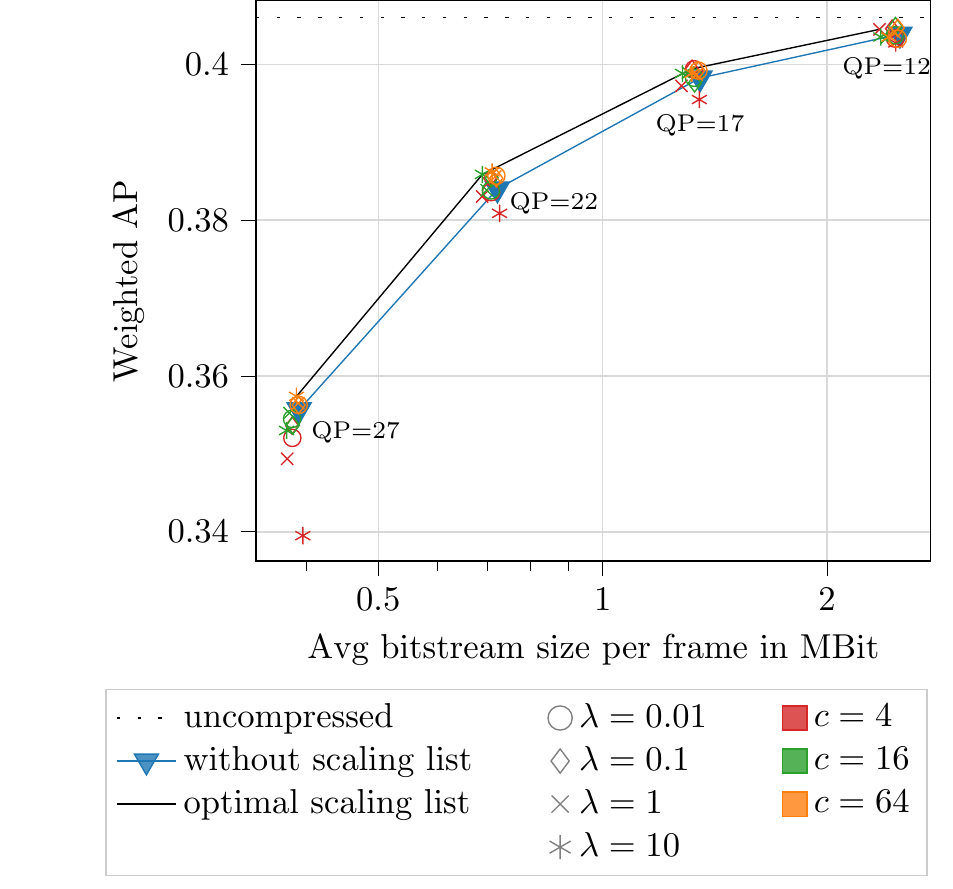}
	\vspace{-3mm}
	\caption{Coding performance of VTM-10.0 depending on the used scaling list for the Cityscapes validation set and Mask R-CNN as analysis network. The black dotted line corresponds to the wAP on uncompressed input data. The marker shape and color denote the selected $\lambda$ and $c$ values, respectively, when training the scaling lists with the proposed method.}
	\label{fig:plot results}
\end{figure}

\subsection{Influence of Training Parametrization on Coding Performance}
As shown in Section~\ref{sec: Proposed Scaling List Generation}, the trained scaling matrices are highly influenced by the initial choice of the noise strength $c$ and the loss weighting $\lambda$. In order to find the best configuration in terms of coding gains, we measured the coding performance of VTM-10.0 with each scaling list derived from the twelve combinations shown in Fig.~\ref{fig:resulting scaling lists}. The resulting rate-wAP points are depicted in Fig.~\ref{fig:plot results}. In most cases, the resulting points from the scaling lists show a lower rate or a higher wAP than VTM without frequency-specific quantization. The BDR-values listed in Table~\ref{tab: BDR values different scaling lists} confirm this observation that VTM coding with the proposed VCM-optimized scaling lists results in bitrate savings over VTM coding without scaling lists for nearly all investigated combinations. At best, the coding gains achieve 7.7\,\% of bitrate savings for $c=16$ and $\lambda=10$.  

Another observation from Fig.~\ref{fig:plot results} is that the optimal scaling list depends on the selected $\QP$. As a rule of thumb for choosing $c$, it can be said that the noise strength $c$ in training shall roughly be similar to the quantization step size $\Delta$ defined via the $\QP$ as in \eqref{eq:qp to delta}. This effect is especially observable for a $\QP$ of 27, where the points for $c=4$ result in a worse coding performance than for $c=64$. Such a relationship cannot be found for $\lambda$. There, our experiments suggest testing multiple combinations for practical applications to find the best scaling list for each $\QP$. By doing so, we obtain the hand-optimized scaling list resulting in the black curve in Fig.~\ref{fig:plot results} and BDR savings of 8.9\,\% over VTM-10.0 without scaling lists.


\begin{table}
	\centering
	\caption{BDR values in \% for the measurement points in Fig.~\ref{fig:plot results}. VTM without scaling list is taken as anchor. Negative values denote bitrate savings over the anchor. The markers correspond to the markers used in Fig.~\ref{fig:plot results}.}
	\begin{tabular}{rr|p{0.0mm}rp{0.0mm}rp{0.0mm}r}
		\toprule
		&      &                                                           \multicolumn{6}{c}{$c$}                                                            \\
		&      &           \multicolumn{2}{c}{4}            &             \multicolumn{2}{c}{16}             &             \multicolumn{2}{c}{64}             \\ \midrule
		\multirow{4}{*}{\tabrotate{$\lambda$}} & 0.01 & \showmark{crimson2143940}{o}        & -1.4 & \showmark{forestgreen4416044}{o}        & -1.3 & \showmark{darkorange25512714}{o}        & -4.7 \\
		&  0.1 & \showmark{crimson2143940}{diamond}  & -6.0 & \showmark{forestgreen4416044}{diamond}  & -3.5 & \showmark{darkorange25512714}{diamond}  & -4.6 \\
		&    1 & \showmark{crimson2143940}{x}        & -0.6 & \showmark{forestgreen4416044}{x}        & -3.4 & \showmark{darkorange25512714}{x}        & -4.6 \\
		&   10 & \showmark{crimson2143940}{asterisk} & 15.1 & \showmark{forestgreen4416044}{asterisk} & -7.7 & \showmark{darkorange25512714}{asterisk} & -7.0 \\ \bottomrule
	\end{tabular}
	\label{tab: BDR values different scaling lists}
\end{table}

\begin{table}
	\centering
	\caption{BDR values in \% for Cityscapes validation images for VTM with the corresponding scaling list for the three quality metrics PSNR, VMAF, and wAP. VTM without scaling list is taken as anchor.}
	\begin{tabular}{l|l|rrr}
		\toprule
		\makecell[tl]{\\Codec} & \makecell[tl]{\\Scaling list} & \makecell[t]{BDR\\PSNR} & \makecell[t]{BDR \\VMAF} &  \makecell[t]{BDR\\wAP} \\
		\midrule
		VTM & JPEG-like~\cite{deLagrange2019} & 8.3 & 2.6 & -4.4 \\
		VTM & HEVC-default~\cite{deLagrange2019} & 1.5 & 0.0 & -4.2\\
		VTM & optimal scaling list & 5.4 & 5.9 & -8.9 \\
		\bottomrule
	\end{tabular}
	\label{tab:comparison scaling lists}
\end{table}


\subsection{Comparison with HVS-optimized Scaling Lists}
The default HVS-optimized scaling lists for JPEG and HEVC also increase the quantization step size for large DCT-frequency coefficients similar to our proposed scaling lists. Thus, we also measure their VCM coding performance as comparison. Table~\ref{tab:comparison scaling lists} shows the BDR values for the quality metrics of PSNR, VMAF, and wAP for two HVS-based and our proposed optimal scaling list. Both scaling lists were taken from~\cite{deLagrange2019}. The BDR values show that the HVS-optimized scaling lists result in bitrate savings around 4\,\% for the investigated VCM-scenario because they also favor low frequencies. However, due to its large optimization towards the final analysis network, encoding the Cityscapes dataset with our proposed scaling lists results in more than twice the bitrate savings. The large coding gains for the VCM scenario come with coding losses for the classic coding scenarios measured by PSNR and VMAF for all three scaling lists.
\vspace{-1mm}
\section{Conclusion}
\vspace{-2mm}
In this paper, we proposed a novel data-driven method to obtain trained scaling lists that are optimized for VVC intra coding in M2M scenarios. Our measurements revealed that the applied Mask R-CNN network is less sensitive to high DCT frequencies similar to the HVS. Thus, a larger quantization step size was assigned to those coefficients. By that, our standard-compliant optimization results in up to 7.7\,\% of bitrate savings over conventional VVC coding with static quantization intervals. When selecting the optimal scaling list for each $\QP$ by hand, we were able to further increase the coding gains to 8.9\,\%. 
Future research has to show, whether the proposed method also achieves similar coding gains for video coding with low-delay P or randomaccess.
Furthermore, we plan to evaluate whether our proposed method also results in coding gains for other analysis networks and tasks.
Moreover, we believe that our method could also be adapted for the HVS by substituting $\taskLoss$ with a suitable metric representing HVS.

\newpage
\bibliographystyle{IEEEbib}
\setstretch{0.01}
\bibliography{/home/fischer/Paper/literature_M2M_communication.bib,/home/fischer/Paper/jabref_used_software.bib}

\end{document}